\documentclass{article}

\usepackage{graphicx}
\usepackage{amssymb}
\usepackage{latexsym}
\usepackage{graphicx}
\usepackage{epstopdf}
\usepackage{caption}
\usepackage{url}
\usepackage{xcolor}
\usepackage{amsmath}
\usepackage{color, colortbl}
\usepackage{enumitem}

 \usepackage{here}
\usepackage{lineno}

\usepackage{tabularx}
\usepackage{array}

\usepackage{multirow}

\usepackage{PRIMEarxiv}

\usepackage[utf8]{inputenc} % allow utf-8 input
\usepackage[T1]{fontenc}    % use 8-bit T1 fonts
\usepackage{hyperref}       % hyperlinks
\usepackage{url}            % simple URL typesetting
\usepackage{booktabs}       % professional-quality tables
\usepackage{amsfonts}       % blackboard math symbols
\usepackage{nicefrac}       % compact symbols for 1/2, etc.
\usepackage{microtype}      % microtypography
\usepackage{lipsum}
\usepackage{fancyhdr}       % header
\usepackage{graphicx}       % graphics
\graphicspath{{media/}}     % organize your images and other figures under media/ folder
\usepackage{makecell}
\title{Sentiment Analysis of Citations in Scientific Articles Using ChatGPT: Identifying Potential Biases and Conflicts of Interest}

\author{
  Walid Hariri \\
  Labged Laboratory, Computer Science department\\ Badji Mokhtar Annaba University, Algeria\\
  walid.hariri@univ-annaba.dz}

\begin{document}
\maketitle

\begin{abstract}
Scientific articles play a crucial role in advancing knowledge and informing research directions. One key aspect of evaluating scientific articles is the analysis of citations, which provides insights into the impact and reception of the cited works. This article introduces the innovative use of large language models, particularly ChatGPT, for comprehensive sentiment analysis of citations within scientific articles. By leveraging advanced natural language processing (NLP) techniques, ChatGPT can discern the nuanced positivity or negativity of citations, offering insights into the reception and impact of cited works. Furthermore, ChatGPT's capabilities extend to detecting potential biases and conflicts of interest in citations, enhancing the objectivity and reliability of scientific literature evaluation. This study showcases the transformative potential of artificial intelligence (AI)-powered tools in enhancing citation analysis and promoting integrity in scholarly research.
\end{abstract}

% keywords can be removed

\section{Introduction}
\label{sec:intro}
\begin{figure}[h]
\centering
\includegraphics[width=0.96\textwidth]{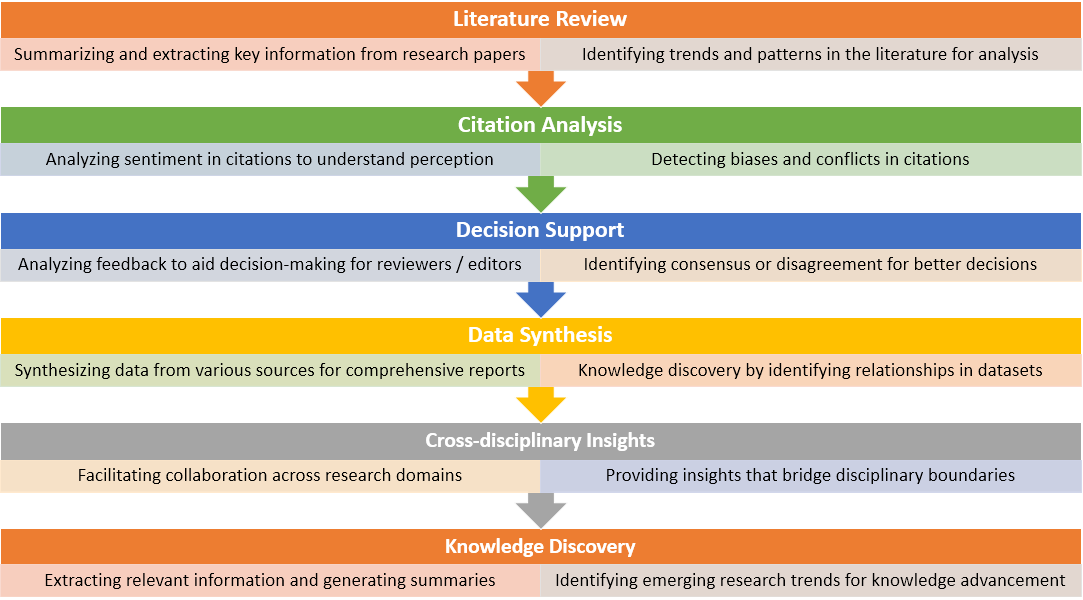}
\caption{Applications of ChatGPT in scientific research.}
\label{fig:fields}
\end{figure}

Scientific publications are instrumental in advancing knowledge and shaping research efforts \cite{huang2023role}. ChatGPT has significantly transformed the landscape of scientific research by offering a multitude of tools and capabilities for writing assistant \cite{imran2023analyzing}, peer-reviewing \cite{mehta2024application}, literature review \cite{haman2023using}, translation \cite{hariri2023unlocking}, writing evaluation \cite{steiss2024comparing,zaabi2023review}, and more. Figure \ref{fig:fields} provides a glimpse into various fields where ChatGPT can be effectively utilized in scientific research.

Sentiment analysis, a field that utilizes natural language processing and machine learning techniques boasts a wide array of applications \cite{hasan2019sentiment}. These include event prediction, examination of public sentiment regarding social issues, and comprehension of customer satisfaction levels and attitudes towards products or services provided by a company. However, analyzing sentiment from scientific article citations is a very difficult task using existing NLP and machine learning techniques especially in complex data like tweets and scientific citations \cite{william2023natural,tubishat2023sentiment}.

Sentiment analysis in scientific citations has gained attention due to increased publication availability \cite{yousif2019survey}. While scholarly databases offer valuable citation information for researchers to exchange ideas, existing metrics across various scientific databases such as Google Scholar, Scopus, Semantic Scholar and Web of Science fail to capture the nuanced nature of citations, which can range from positive to negative. Hence, understanding citation sentiment in scientific articles is crucial for several reasons. Firstly, it provides insights into how the scientific community perceives and evaluates research findings even post-publication. For instance, a citation that praises the methodology or results of a study indicates positive sentiment, reflecting recognition and validation from peers. In contrast, a citation that criticizes certain aspects of a study suggests negative sentiment, highlighting areas of controversy or divergence in opinions.

Secondly, citation sentiment analysis helps researchers gauge the impact and influence of their work \cite{susnjak2024applying}. Positive citations from reputable journals or renowned researchers can elevate the visibility and credibility of a study, attracting more attention and potential collaborations. Conversely, negative citations may prompt researchers to re-evaluate their methodologies or interpretations, leading to improvements in future research directions. Furthermore, comprehending citation sentiment is essential for reviewers and editors during the peer-review process, particularly when handling already published submissions like preprints, or to identify biased citations or potential conflicts of interest. Positive citations can reinforce the strengths and contributions of a manuscript, supporting its acceptance and publication. Conversely, negative citations may raise concerns about the robustness or validity of the findings, prompting reviewers to scrutinize the study more closely before making decisions.

Accordingly, citation sentiment analysis goes beyond mere acknowledgment of sources; it serves as a barometer of scholarly impact, consensus, and critique within the scientific community. An additional area where citation sentiment analysis proves valuable within scientific research lies in its ability to identify non-reproducible studies, a growing concern among researchers \cite{xu2015citation}. This understanding contributes significantly to the advancement of scientific knowledge. Notably, a paper published in a highly-ranked journal with a negative sentiment analysis ranking may be less engaging than one published in a lower-ranked journal with a highly positive sentiment analysis of citations.

The rest of this article is structured as follows: Section 2 presents challenges in citation analysis in the scientific research field. In section 3, we present ChatGPT as a tool for sentiment analysis. We then present analytical approaches for positive and negative citations. Identifying biases and conflicts of interest are explained in Section 5, where the application of ChatGPT in reviewing and research is highlighted in Section 6. Section 7 delves into its application in the editor's decision. Limitations and future directions are given in Section 8. Conclusions end the paper.

\section{Challenges in citation analysis}
Citation analysis in scientific articles presents several challenges that researchers and reviewers must navigate. One of the primary challenges is the diversity of citation sentiments. Citations can range from highly positive endorsements to neutral acknowledgments and even negative criticisms. For example, a positive citation may commend a study's innovative approach or robust methodology, while a negative citation may question the validity of certain findings or raise concerns about methodological flaws.

Another challenge is the subjectivity inherent in interpreting citation sentiments. What one researcher perceives as a positive endorsement, another may view as a neutral reference. For instance, a citation that simply mentions a study without providing context or evaluation may be interpreted differently by different readers. This subjectivity can lead to discrepancies in how citations are analyzed and understood. Furthermore, the sheer volume of citations in scientific literature poses a challenge in comprehensive analysis. Research articles often cite numerous sources to support their arguments or provide background information. Reviewers and researchers may struggle to assess the sentiment of each citation accurately, especially in large-scale studies with extensive reference lists.

Moreover, the evolving nature of scientific knowledge and perspectives adds complexity to citation analysis. Over time, the perception of a study may change, leading to shifts in how its citations are interpreted. For example, a study that was initially praised for its groundbreaking findings may face criticisms or reassessments in light of new evidence or developments in the field. This is especially true for older citations, where a positive citation about a work may become negative over time as newer research surpasses its findings, rendering it less important or impactful in the current scientific landscape.

Additionally, the context in which citations are used can influence their perceived sentiment. Citations within the introduction or discussion sections of an article may carry different implications than those within the methodology or results sections. Understanding this contextual variability is essential for accurate citation sentiment analysis.

Ultimately, challenges in citation analysis stem from the diversity, subjectivity, volume, evolution, and contextuality of citations in scientific literature. Overcoming these challenges requires careful consideration, nuanced interpretation, and advanced analytical tools such as ChatGPT to assist in navigating the complexities of citation sentiment analysis.

\section{ChatGPT: a tool for sentiment analysis}
Exploring how ChatGPT can be leveraged to analyse citation sentiments efficiently delves into the structural prowess of large language models like transformers. These models, such as ChatGPT, are built on transformer architectures that excel in processing and understanding natural language text. The transformer architecture's attention mechanisms allow ChatGPT to capture intricate relationships between words and phrases, enabling nuanced sentiment analysis within citations.

ChatGPT's ability to comprehend context and context shifts is another key factor in its efficient analysis of citation sentiments. By considering the surrounding text and the citation's placement within an article, ChatGPT can discern the tone, intent, and underlying sentiment with remarkable accuracy. This contextual understanding ensures that the sentiment analysis reflects the nuanced nuances present in scholarly citations.

Furthermore, ChatGPT's training on vast amounts of scientific literature enables it to recognise domain-specific expressions, keywords, and patterns associated with positive or negative sentiments in citations. This extensive training enhances ChatGPT's proficiency in analysing citation sentiments across diverse research fields and topics.

\begin{figure}[H]
\centering
\includegraphics[width=0.80\textwidth]{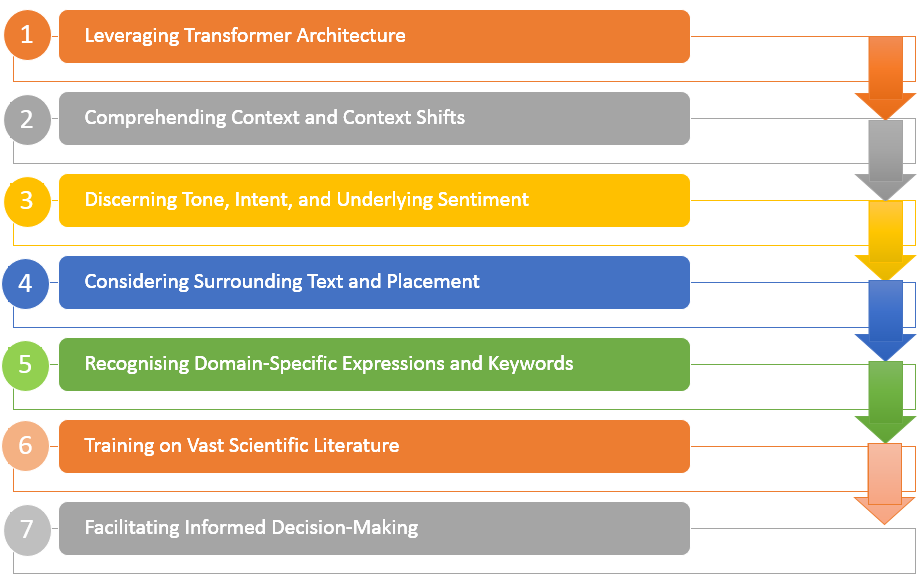}
\caption{Sentiment analysis process using chatGPT.}
\label{fig:process}
\end{figure}

By taking into account the capabilities of ChatGPT to analyse sentiment, leveraging ChatGPT's transformer-based architecture and contextual understanding allows for efficient and accurate analysis of citation sentiments. This structural prowess, combined with domain-specific training, empowers ChatGPT to handle large volumes of citations with speed, precision, and scalability, facilitating informed decision-making and deeper insights into scholarly impact. Figure \ref{fig:process} presents the process of sentiment analysis using ChatGPT. 

\section{Analytical approaches for positive and negative citations}

Analyzing positive and negative citations involves a multifaceted approach, combining keywords and expressions to accurately gauge sentiment. Keywords such as "groundbreaking," "significant contribution," and "remarkable findings" often signify positive citations, while terms like "limitations," "inconsistencies," and "weak experimental design" tend to indicate negative sentiments. Additionally, expressions like "revolutionary study" and "exceptional methodology" further reinforce positivity, while phrases such as "lack of statistical rigor" and "limited scope" contribute to a negative assessment. By considering both keywords and expressions, researchers can effectively determine the positivity or negativity of citations in scientific articles.

Below are examples of sentences commonly found in articles along with their sentiment analysis:
\begin{itemize}
    \item \textbf{Positive citations:}
"The groundbreaking study by Smith et al. (2020) significantly advances our understanding of climate change impacts."
"Jones' meticulous research methodology provides a solid foundation for future investigations in the field."

\item \textbf{Negative citations:}
"Contrary to the claims made by Johnson et al. (2019), subsequent studies have shown inconsistencies in their experimental design."
"The limitations of Brown's study highlight the need for more comprehensive data collection methods in similar research."
\end{itemize}

\begin{figure}[h]
\centering
\includegraphics[width=0.99\textwidth]{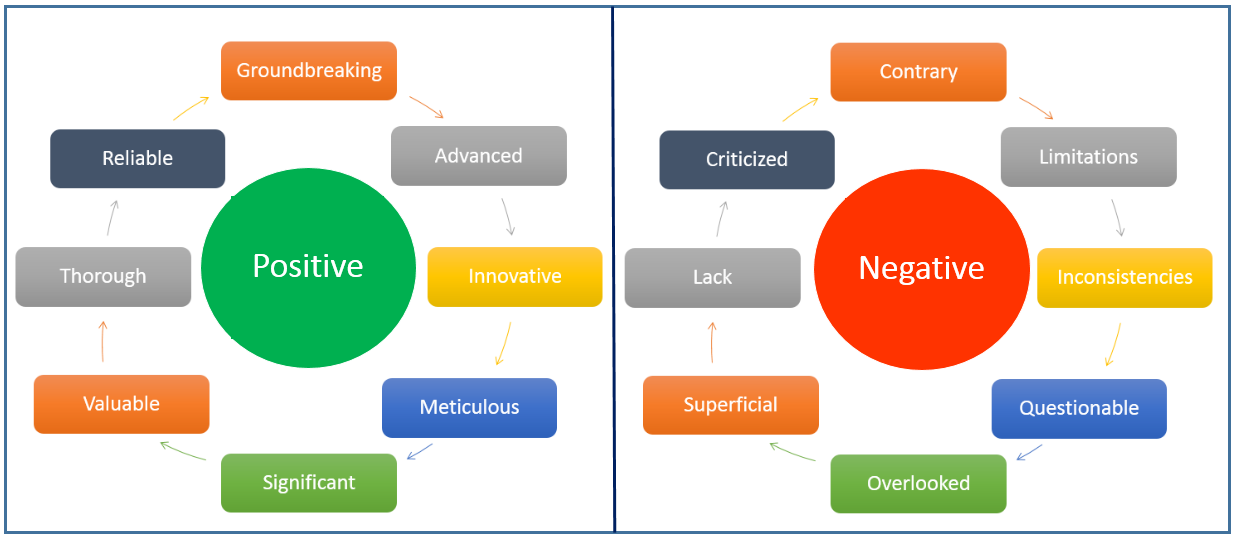}
\caption{Keywords for determining positive and negative sentiments in citations.}
\label{fig:words}
\end{figure}

These examples demonstrate the contrasting sentiments expressed in citations, ranging from praise and endorsement to critique and reservation. Figure \ref{fig:words} depicts the most used words for positive and negative citations.
Therefore, conducting a thorough sentiment analysis of citations could be highly significant even after publication as it offers insights into the researcher's impact within their research community..  

Traditional metrics, however, like the total number of citations in scholarly databases (such as Google Scholar, Scopus, and Web of Science) and the h-index have limitations as they fail to provide insights into the nature of the citation \cite{costas2007h}. It's important to consider additional features when analyzing sentiment in citations, especially given the challenge of data balancing, as demonstrated in \cite{karim2022comprehension}. The experiments in this study explore accuracy with imbalanced datasets and assess the impact of sampling techniques and feature types.

Each field within scholarly writing can exhibit its unique keywords, citation structure, and manner of expression. For example, in Computer Science, words like "algorithm efficiency" and "code quality" might signal positive sentiments, while phrases like "security vulnerabilities" and "error handling" could denote negativity. Similarly, in Medicine, terms such as "promising results" and "population health implications" may indicate positive viewpoints, whereas "methodology flaws" and "high false-positive rates" might imply negativity. 
Table \ref{tab:citations} summarises the examples of positive and negative citations, showcasing the different sentiments conveyed in scholarly writing in many domains including computer science, architecture, physics chimestry, etc.
\begin{table}[H]
\footnotesize
\centering
\caption{Examples of positive and negative citations with domain.}
\begin{tabular}{|l|l|p{10cm}|}
\hline
\textbf{Domain} & \textbf{Sentiment} & \textbf{Example Sentences} \\
\hline
\multirow{4}{*}{Computer Science} & Positive & "The algorithm proposed by Smith et al. (2022) significantly improves efficiency in data processing tasks." \newline "Jones' software development methodology sets a new standard for code quality and reliability." \\
\cline{2-3}
& Negative & "Contrary to the findings of Johnson et al. (2021), recent studies have identified vulnerabilities in their security model." \newline "The limitations of Brown's computational model highlight the need for more robust error handling mechanisms." \\
\hline
\multirow{4}{*}{Architecture} & Positive & "The innovative design approach by Smith and team (2024) showcases a harmonious blend of functionality and aesthetics." \newline "Jones' sustainable architecture principles demonstrate a commitment to environmental conservation." \\
\cline{2-3}
& Negative & "Despite the acclaim, Johnson's architectural design lacks practicality and usability." \newline "Brown's architectural concept has been criticised for its lack of structural integrity." \\
\hline
\multirow{4}{*}{English Literature} & Positive & "Smith's literary analysis sheds new light on the complexities of Shakespearean tragedies." \newline "Jones' critical review of modern poetry enriches our understanding of contemporary literary trends." \\
\cline{2-3}
& Negative & "Contrary to prevailing opinions, Johnson's literary interpretation overlooks crucial thematic elements." \newline "The limitations of Brown's literary critique lie in its superficial examination of character motivations." \\
\hline
\multirow{4}{*}{Physics} & Positive & "The experimental findings of Smith et al. (2020) confirm the existence of quantum entanglement at macroscopic scales." \newline "Jones' theoretical framework provides a novel perspective on the unified theory of particle physics." \\
\cline{2-3}
& Negative & "Johnson's theoretical model fails to account for observed phenomena in quantum mechanics." \newline "The conclusions drawn by Brown et al. (2018) contradict established principles of thermodynamics." \\
\hline
\multirow{4}{*}{Chemistry} & Positive & "The synthesis method developed by Smith and colleagues (2024) offers a sustainable solution for chemical waste reduction." \newline "Jones' research on catalyst efficiency has wide-ranging applications in industrial processes." \\
\cline{2-3}
& Negative & "Despite initial promise, Johnson's chemical reaction mechanism lacks reproducibility in experimental settings." \newline "Brown's chemical analysis overlooks key factors impacting reaction kinetics." \\
\hline
\multirow{4}{*}{Electrical Engineering} & Positive & "The innovative circuit design by Smith et al. (2020) significantly improves power efficiency in electronic devices." \newline "Jones' research on renewable energy sources contributes to advancements in sustainable technologies." \\
\cline{2-3}
& Negative & "Johnson's electrical circuit design is prone to voltage fluctuations under varying load conditions." \newline "The limitations of Brown's electrical system design result in inefficiencies and overheating." \\
\hline
\multirow{4}{*}{Biology} & Positive & "The discovery by Smith's research team (2020) offers new insights into genetic mechanisms underlying disease progression." \newline "Jones' experimental methodology in cell biology sets a benchmark for precision and reproducibility." \\
\cline{2-3}
& Negative & "Contrary to expectations, Johnson's experimental results show no significant impact on cellular function." \newline "The limitations of Brown's biological model restrict its applicability to real-world scenarios." \\
\hline
\multirow{4}{*}{Medicine} & Positive & "The clinical trial led by Smith et al. (2020) demonstrates promising results in cancer treatment efficacy." \newline "Jones' medical research on public health interventions has positive implications for population health." \\
\cline{2-3}
& Negative & "Johnson's clinical study methodology lacks adequate control measures, raising questions about result validity." \newline "Brown's medical diagnosis algorithm exhibits high false-positive rates, impacting diagnostic accuracy." \\
\hline
\end{tabular}
\label{tab:citations}
\end{table}

\section{Identifying biases and conflicts of interest in citations with ChatGPT}
A biased citation refers to a citation that shows a skewed or one-sided perspective, often lacking objectivity or fairness in its evaluation of the cited work \cite{urlings2021citation,ray2024citation}. On the other hand, a conflict of interest in a citation arises when the author or authors have financial \cite{thompson2017understanding}, professional, or personal connections that could influence the objectivity or credibility of the cited information.

Identifying biases and conflicts of interest in citations with ChatGPT involves a multifaceted approach that leverages advanced natural language processing capabilities. ChatGPT can analyse the language, context, author affiliations, and citation patterns to detect potential biases and conflicts of interest in scholarly articles.

Biased self-citation refers to the practice where authors excessively cite their own work in a way that promotes their own research or enhances their academic reputation.
Chorus et al. \cite{chorus2016large} studied the impact of self-citation in academic journals using a specific measure. It is determined by the division of i) the proportion of self-citations within journals to papers published within the last two years, and ii) the proportion of self-citations within journals to papers published in earlier years.

This biased self-citation can manifest in various forms, such as:

\begin{itemize}
    \item \textbf{Enhancing visibility:} Authors may cite their own previous work extensively to increase the visibility and perceived importance of their research within a specific field.

    \item \textbf{Inflating citation metrics:} By citing their own work multiple times, authors can artificially inflate their citation metrics, such as h-index or citation count, which are often used as measures of academic impact.

    \item \textbf{Promoting personal agenda:} Biased self-citation can also be used to promote a particular viewpoint, theory, or methodology that aligns with the author's personal agenda or research interests.

    \item \textbf{Neglecting alternative perspectives:} Over-reliance on self-citation may lead to a neglect of alternative perspectives or relevant research by other scholars, potentially biasing the literature review and overall scholarly discourse.

    \item \textbf{Exaggerating influence:} Authors may use biased self-citation to exaggerate the influence and significance of their own contributions within their field of study.
\end{itemize}

One aspect of this analysis is examining the language and sentiment used in citations. ChatGPT can identify overly positive or negative language that may indicate bias. For example, a citation that excessively praises a study without providing critical analysis or context may reflect a biased viewpoint. Conversely, a citation that consistently criticizes a study without acknowledging its merits may also indicate bias. For instance:
\begin{itemize}
    \item \textbf{Positive bias:} "Smith's groundbreaking study completely revolutionizes the field."
    \item \textbf{Negative bias:} "Johnson's research methodology lacks rigour and validity."
\end{itemize}

Moreover, ChatGPT can scrutinize author affiliations and backgrounds to identify conflicts of interest. Citations from authors with financial or professional ties to a company, organization, or specific research agenda may suggest potential conflicts. Table \ref{tab:conflicts} shows some examples of citations with and without conflicts of interest.

\begin{table}[H]
\centering
\caption{Examples of citations with and without conflicts of interest.}
\begin{tabular}{|p{5cm}|p{10cm}|}
\hline
\textbf{Type} & \textbf{Example Citations} \\
\hline
Conflict of Interest & \textbullet{} According to Jones, sponsored by Company X, the benefits of Product Y are evident. \\
& \textbullet{} Smith's research, funded by Corporation Z, demonstrates the superiority of their product over competitors. \\
& \textbullet{} The study by Johnson et al., affiliated with Pharmaceutical Company A, promotes the use of Drug B as the preferred treatment option. \\
\hline
Neutral (No Conflict of Interest) &\textbullet{} In Smith’s study, the methodology employed was thorough and well-documented, providing valuable insights into the topic. \\
&\textbullet{} Jones' research, although funded by various sources, presents unbiased findings that contribute significantly to the field. \\
&\textbullet{} The study by Brown et al., despite industry collaborations, maintains scientific integrity in its analysis and conclusions. \\
\hline
\end{tabular}
\label{tab:conflicts}
\end{table}

Furthermore, ChatGPT can compare citation sentiments across different articles and authors to detect patterns of bias or conflict. ChatGPT can also detect AI-written articles where fictitious references can be identified \cite{ariyaratne2023comparison}. By considering the context of citations within the citing article, ChatGPT can determine whether citations are used objectively to support arguments or subjectively to promote certain viewpoints. This contextual analysis helps uncover subtle biases or conflicts that may not be immediately apparent.

Moreover, ChatGPT can analyse citation patterns within a research field to identify clusters of biased or conflicted citations. For example, if a particular study receives overwhelmingly positive citations from authors affiliated with a specific industry or institution, it may suggest a biased narrative or conflict of interest. Overall, ChatGPT's comprehensive NLP capabilities enable it to uncover biases and conflicts of interest in citations, contributing to the integrity and credibility of scholarly research. 

Table \ref{tab:biases} illustrates examples of biased and unbiased citations in scientific articles. Biased citations often favor specific theories or products, neglecting alternatives and limitations, and are influenced by personal beliefs or hidden agendas. In contrast, unbiased citations provide a balanced evaluation of different approaches, objectively presenting strengths and weaknesses, and prioritize transparent, data-driven results over biases.

\begin{table}[H]
\centering
\caption{Examples of citations with and without biases in scientific articles.}
\begin{tabular}{|p{5cm}|p{10cm}|}
\hline
\textbf{Type} & \textbf{Example Citations} \\
\hline
Biased & 
\textbullet{} According to Smith, who favors Theory X, all evidence points to its superiority over competing theories. \\
& \textbullet{} Johnson’s study, influenced by personal beliefs, heavily promotes the benefits of Method Y without considering alternative approaches. \\
& \textbullet{} The research by Brown et al., with a clear bias towards Product Z, overlooks potential drawbacks and limitations. \\
& \textbullet{} White’s analysis, driven by a specific agenda, downplays contradictory findings to support a predetermined conclusion. \\
& \textbullet{} In the article by Green, the author’s affiliation with Company A leads to a one-sided portrayal of the effectiveness of their product. \\
\hline
Unbiased & 
\textbullet{} In Jones' study, a comprehensive review of multiple theories is presented, highlighting strengths and weaknesses objectively. \\
& \textbullet{} The research by Johnson et al., while exploring various methods, provides a balanced assessment of each approach without favoritism. \\
& \textbullet{} Brown's analysis, despite personal preferences, critically evaluates different perspectives and acknowledges nuances in the data. \\
& \textbullet{} White's investigation, conducted without preconceived notions, carefully considers all available evidence before drawing conclusions. \\
& \textbullet{} The study by Green and colleagues, with a focus on data-driven analysis, avoids bias by transparently reporting methodology and results. \\
\hline
\end{tabular}
\label{tab:biases}
\end{table}

\section{Applications in reviewing and research}
ChatGPT's capabilities can assist journal reviewers or conference committee members by automating tasks such as summarising articles, extracting key findings, and identifying potential biases in citations. Moreover, it can aid researchers in conducting literature reviews by analysing sentiment patterns and identifying prevailing opinions or controversies within a field. Additionally, ChatGPT's ability to analyse citation sentiments contributes to more objective and transparent research analyses. By leveraging ChatGPT's NLP capabilities, reviewers and researchers can navigate the scholarly landscape more efficiently, leading to more informed decision-making and robust research outcomes.

Recently, a growing number of papers are first released as preprints on various repositories such as arXiv, ResearchSquare, medRxiv, bioRxiv, ChemRxiv, among others. The goal of preprints is to disseminate research in open access, speed up, and enhance the peer review process \cite{stojanovski2024preprints}.

Preprints often garner significant attention and citations even before undergoing peer review and formal publication in a journal. Notably, many journals now recognise and accept preprints as valid publications, considering them as officially published works upon submission. This trend has led to preprints being an essential part of scholarly communication, offering insights into researchers' satisfaction within their respective fields even before the peer review process begins. Analyzing preprint citations can serve as a valuable indicator of the reception and impact of research, providing valuable context for reviewers and editors during manuscript evaluation.

\section{Applications in editorial decision-making with ChatGPT}
Editorial decision-making plays a critical role in the publication process, influencing the quality and impact of scholarly works. ChatGPT offers a range of applications that enhance and streamline the decision-making process for journal editors. Here are some key ways in which ChatGPT can be applied in editorial decision-making:

\begin{itemize}
    \item \textbf{Sentiment analysis:} ChatGPT can analyse the sentiment of each reviewer's feedback, identifying positive and negative sentiments. This allows editors to gauge the overall sentiment of the feedback and understand the reviewers' perspectives.

    \item \textbf{Contextual analysis:} ChatGPT can consider the context of the feedback within the article and the specific points raised by each reviewer. This contextual analysis helps editors evaluate the validity and relevance of the feedback in relation to the article's content.

    \item \textbf{Consistency checking:} Editors can use ChatGPT to check for consistency in feedback across different sections of the article. Inconsistencies or conflicting feedback can be flagged for further review and discussion.

    \item \textbf{Objective decision-making:} ChatGPT provides editors with an objective tool to supplement their decision-making process. By analysing the sentiments and content of reviewers' feedback, editors can make more informed and impartial decisions.

    \item \textbf{Identifying key points:} ChatGPT can highlight key points or areas of contention in the feedback, helping editors focus on crucial aspects that require attention or clarification.

    \item \textbf{Enhancing communication:} ChatGPT can facilitate communication between editors and reviewers by providing a structured analysis of feedback. This promotes constructive dialogue and consensus-building.

    \item \textbf{Feedback integration:} Editors can integrate ChatGPT's analysis with their own expertise and judgement, synthesising the insights gained from both human and AI-driven analyses to arrive at a well-rounded decision.

    \item \textbf{Plagiarism detection:}  ChatGPT can assist in plagiarism detection by comparing the manuscript with existing literature and identifying similarities or copied content. This helps editors ensure the originality and integrity of submitted manuscripts, maintaining the credibility of the peer review process.
\end{itemize}

By fulfilling the aforementioned tasks, utilising ChatGPT in editor decision-making enhances the efficiency, objectivity, and transparency of the peer review process, ultimately enhancing the quality and credibility of published research.

\section{Limitations and future directions}
While ChatGPT offers significant potential in enhancing various aspects of academic publishing, it is important to acknowledge its limitations and explore future directions for improvement and innovation.

\begin{itemize}
    \item \textbf{Model limitations:} ChatGPT's performance in sentiment analysis may be impacted by biases in the training data and the model's reliance on pre-existing knowledge. Future research should focus on mitigating these biases and improving the model's ability to handle diverse contexts and terminology.

    \item \textbf{Contextual understanding:} One of the challenges faced by ChatGPT is its limited ability to grasp nuanced context, leading to potential misinterpretations in sentiment analysis. Future developments should aim to enhance ChatGPT's contextual understanding, especially in domain-specific language and academic discourse.

    \item \textbf{Ethical considerations:} The integration of AI tools like ChatGPT in decision-making processes raises ethical considerations, including transparency, accountability, and potential biases. Future research should address these ethical concerns and establish guidelines for responsible AI usage in academic publishing.

    \item \textbf{Human-AI collaboration:} There is a need to foster better collaboration between human editors/reviewers, conference committee members, and AI tools like ChatGPT. Future directions should focus on developing AI systems that augment human expertise rather than replacing it, promoting a symbiotic relationship between humans and AI.

    \item \textbf{Continuous improvement:} ChatGPT requires continuous training and fine-tuning to improve its accuracy and reliability in sentiment analysis and decision-making tasks. Future efforts should prioritise ongoing model refinement and adaptation to evolving research needs.

    \item \textbf{Interdisciplinary applications:} Beyond paper reviewing and research analysis, ChatGPT has the potential for interdisciplinary applications such as data synthesis, knowledge discovery, medical \cite{hariri2023analyzing,haouli2023exploring}, and cross-disciplinary research insights \cite{kadhim2022face}. Future research should explore these diverse applications to unlock ChatGPT's full potential.

    \item \textbf{User interface and accessibility:} Designing user-friendly interfaces that integrate ChatGPT seamlessly into existing editorial workflows is crucial for enhancing accessibility and usability for editors, reviewers, and conference committee members.

    \item \textbf{Data privacy and security:} Ensuring data privacy and security in AI-driven processes like ChatGPT is paramount. Future research should focus on implementing robust data protection measures, complying with regulatory requirements, and safeguarding sensitive information.

    \item \textbf{Collaborative research:} Collaborative research efforts are essential to explore the broader implications of AI-driven technologies in scholarly communication. Future collaborations should foster interdisciplinary dialogue and knowledge sharing among reviewers, editors, conference committee members, and AI experts.

    \item \textbf{Long-term impact assessment:} Continuous monitoring and assessment of ChatGPT's long-term impact on academic publishing are necessary. Future research should develop mechanisms for evaluating outcomes, addressing feedback, and iteratively improving AI systems for sustained effectiveness.
\end{itemize}

By addressing these limitations and exploring future directions, the integration of ChatGPT and similar AI technologies can lead to transformative advancements in academic publishing, enhancing efficiency, objectivity, and transparency in decision-making processes for reviewers, journal editors, conference committee members, and the scholarly community.
\section{Experiment}
In this section, we evaluate the efficiency of ChatGPT in analyzing the sentiment within citations. We collected 50 citations from published scientific papers in journals and proceedings and annotated each citation as neutral, positive, or negative. Subsequently, we used ChatGPT with a specific prompt to classify each citation. The results are presented in Table 4. From this table, we can see that ChatGPT has corrected classified ... out of 75 citations.

\begin{table}[h!]
\centering
\caption{Dataset of citation analysis, and accuracy of ChatGPT classification.}
\begin{tabular}{|c|c|c|c|}
\hline
Category                   & Count& Example&Accuracy \\
\hline
Total number of citations  & 75  & & \\
\hline
Positive                   & 25 & &  \\
\hline
Negative                   & 25 &  & \\
\hline
Neutral                    & 25 &  & \\
\hline
\end{tabular}

\label{tab:example}
\end{table}

\section{Conclusion}
In conclusion, ChatGPT represents a powerful tool with immense potential to revolutionize various aspects of academic publishing. Its ability to analyze sentiment in citations, assist in decision-making for reviewers and editors, and contribute to a more transparent and efficient scholarly communication process is evident. Citation count doesn't accurately reflect a paper's quality because citations can also highlight weaknesses. Therefore, analyzing citation sentiment is crucial for evaluating article quality, impact, and ranking. However, it is crucial to acknowledge the limitations and challenges that come with integrating AI technologies like ChatGPT into academic workflows. Addressing issues such as bias mitigation, contextual understanding, ethical considerations, and user interface design will be essential for maximizing the benefits of ChatGPT while ensuring responsible and equitable usage. Looking ahead, collaborative efforts between researchers, AI developers, reviewers, editors, and conference committee members will play a pivotal role in advancing AI-driven solutions in academic publishing. With continuous improvement, interdisciplinary collaboration, and a commitment to ethical AI practices, ChatGPT and similar technologies have the potential to foster a more objective, efficient, and inclusive scholarly ecosystem.
%Bibliography
\bibliographystyle{unsrt}  
\bibliography{references}

@article{huang2023role,
  title={The role of ChatGPT in scientific communication: writing better scientific review articles},
  author={Huang, Jingshan and Tan, Ming},
  journal={American journal of cancer research},
  volume={13},
  number={4},
  pages={1148},
  year={2023},
  publisher={e-Century Publishing Corporation}
}

@incollection{susnjak2024applying,
  title={Applying bert and chatgpt for sentiment analysis of lyme disease in scientific literature},
  author={Susnjak, Teo},
  booktitle={Borrelia burgdorferi: Methods and Protocols},
  pages={173--183},
  year={2024},
  publisher={Springer}
}

@inproceedings{tubishat2023sentiment,
  title={Sentiment analysis of using chatgpt in education},
  author={Tubishat, Mohammad and Al-Obeidat, Feras and Shuhaiber, Ahmed},
  booktitle={2023 International Conference on Smart Applications, Communications and Networking (SmartNets)},
  pages={1--7},
  year={2023},
  organization={IEEE}
}

@article{ariyaratne2023comparison,
  title={A comparison of ChatGPT-generated articles with human-written articles},
  author={Ariyaratne, Sisith and Iyengar, Karthikeyan P and Nischal, Neha and Chitti Babu, Naparla and Botchu, Rajesh},
  journal={Skeletal radiology},
  volume={52},
  number={9},
  pages={1755--1758},
  year={2023},
  publisher={Springer}
}

@misc{mehta2024application,
  title={The application of ChatGPT in the peer-reviewing process},
  author={Mehta, Vini and Mathur, Ankita and Anjali, AK and Fiorillo, Luca},
  journal={Oral Oncology Reports},
  pages={100227},
  year={2024},
  publisher={Elsevier}
}

@article{haman2023using,
  title={Using ChatGPT to conduct a literature review},
  author={Haman, Michael and {\v{S}}koln{\'\i}k, Milan},
  journal={Accountability in research},
  pages={1--3},
  year={2023},
  publisher={Taylor \& Francis}
}

@article{steiss2024comparing,
  title={Comparing the quality of human and ChatGPT feedback of students’ writing},
  author={Steiss, Jacob and Tate, Tamara and Graham, Steve and Cruz, Jazmin and Hebert, Michael and Wang, Jiali and Moon, Youngsun and Tseng, Waverly and Warschauer, Mark and Olson, Carol Booth},
  journal={Learning and Instruction},
  volume={91},
  pages={101894},
  year={2024},
  publisher={Elsevier}
}

@incollection{thompson2017understanding,
  title={Understanding financial conflicts of interest},
  author={Thompson, Dennis F},
  booktitle={Research Ethics},
  pages={505--508},
  year={2017},
  publisher={Routledge}
}

@article{urlings2021citation,
  title={Citation bias and other determinants of citation in biomedical research: findings from six citation networks},
  author={Urlings, Miriam JE and Duyx, Bram and Swaen, Gerard MH and Bouter, Lex M and Zeegers, Maurice P},
  journal={Journal of Clinical Epidemiology},
  volume={132},
  pages={71--78},
  year={2021},
  publisher={Elsevier}
}

@article{chorus2016large,
  title={A large-scale analysis of impact factor biased journal self-citations},
  author={Chorus, Caspar and Waltman, Ludo},
  journal={PLoS One},
  volume={11},
  number={8},
  pages={e0161021},
  year={2016},
  publisher={Public Library of Science San Francisco, CA USA}
}

@article{ray2024citation,
  title={Citation bias, diversity, and ethics},
  author={Ray, Keisha S and Zurn, Perry and Dworkin, Jordan D and Bassett, Dani S and Resnik, David B},
  journal={Accountability in Research},
  volume={31},
  number={2},
  pages={158--172},
  year={2024},
  publisher={Taylor \& Francis}
}

@article{stojanovski2024preprints,
  title={Preprints Are Here to Stay: Is That Good for Science?},
  author={Stojanovski, Jadranka and Maru{\v{s}}i{\'c}, Ana},
  journal={Second Handbook of Academic Integrity},
  pages={1383--1401},
  year={2024},
  publisher={Springer}
}

@article{yousif2019survey,
  title={A survey on sentiment analysis of scientific citations},
  author={Yousif, Abdallah and Niu, Zhendong and Tarus, John K and Ahmad, Arshad},
  journal={Artificial Intelligence Review},
  volume={52},
  pages={1805--1838},
  year={2019},
  publisher={Springer}
}

@article{hariri2023unlocking,
  title={Unlocking the potential of ChatGPT: A comprehensive exploration of its applications, advantages, limitations, and future directions in natural language processing},
  author={Hariri, Walid},
  journal={arXiv preprint arXiv:2304.02017},
  year={2023}
}

@inproceedings{hasan2019sentiment,
  title={Sentiment analysis with NLP on Twitter data},
  author={Hasan, Md Rakibul and Maliha, Maisha and Arifuzzaman, M},
  booktitle={2019 international conference on computer, communication, chemical, materials and electronic engineering (IC4ME2)},
  pages={1--4},
  year={2019},
  organization={IEEE}
}

@article{imran2023analyzing,
  title={Analyzing the role of ChatGPT as a writing assistant at higher education level: A systematic review of the literature},
  author={Imran, Muhammad and Almusharraf, Norah},
  journal={Contemporary Educational Technology},
  volume={15},
  number={4},
  pages={ep464},
  year={2023},
  publisher={Bastas}
}

@incollection{william2023natural,
  title={Natural Language processing implementation for sentiment analysis on tweets},
  author={William, P and Shrivastava, Anurag and Chauhan, Premanand S and Raja, Mudasir and Ojha, Sudhir Baijnath and Kumar, Keshav},
  booktitle={Mobile Radio Communications and 5G Networks: Proceedings of Third MRCN 2022},
  pages={317--327},
  year={2023},
  publisher={Springer}
}

@article{costas2007h,
  title={The h-index: Advantages, limitations and its relation with other bibliometric indicators at the micro level},
  author={Costas, Rodrigo and Bordons, Mar{\'\i}a},
  journal={Journal of informetrics},
  volume={1},
  number={3},
  pages={193--203},
  year={2007},
  publisher={Elsevier}
}

@inproceedings{xu2015citation,
  title={Citation sentiment analysis in clinical trial papers},
  author={Xu, Jun and Zhang, Yaoyun and Wu, Yonghui and Wang, Jingqi and Dong, Xiao and Xu, Hua},
  booktitle={AMIA annual symposium proceedings},
  volume={2015},
  pages={1334},
  year={2015},
  organization={American Medical Informatics Association}
}

@article{karim2022comprehension,
  title={Comprehension of polarity of articles by citation sentiment analysis using TF-IDF and ML classifiers},
  author={Karim, Musarat and Missen, Malik Muhammad Saad and Umer, Muhammad and Fida, Alisha and Mohamed, Abdullah and Ashraf, Imran and others},
  journal={PeerJ Computer Science},
  volume={8},
  pages={e1107},
  year={2022},
  publisher={PeerJ Inc.}
}

@inproceedings{zaabi2023review,
  title={A review study of ChatGPT applications in education},
  author={Zaabi, Marwa and Hariri, Walid and Smaoui, Nadia},
  booktitle={2023 International Conference on Innovations in Intelligent Systems and Applications (INISTA)},
  pages={1--5},
  year={2023},
  organization={IEEE}
}

@article{hariri2023analyzing,
  title={Analyzing the Performance of ChatGPT in Cardiology and Vascular Pathologies},
  author={Hariri, Walid},
  journal={arXiv preprint arXiv:2307.02518},
  year={2023}
}

@inproceedings{haouli2023exploring,
  title={Exploring vision transformers for automated glaucoma disease diagnosis in fundus images},
  author={Haouli, Imed-Eddine and Hariri, Walid and Seridi-Bouchelaghem, Hassina},
  booktitle={2023 International Conference on Decision Aid Sciences and Applications (DASA)},
  pages={520--524},
  year={2023},
  organization={IEEE}
}

@inproceedings{kadhim2022face,
  title={Face recognition in multiple variations using deep learning and convolutional neural networks},
  author={Kadhim, Thair A and Zghal, Nadia Smaoui and Hariri, Walid and Aissa, Dalenda Ben},
  booktitle={2022 IEEE 9th International Conference on Sciences of Electronics, Technologies of Information and Telecommunications (SETIT)},
  pages={305--311},
  year={2022},
  organization={IEEE}
}

\end{document}